\documentclass[pra]{revtex4}
 \usepackage{amssymb} \usepackage{graphicx}

\begin{document}
 \title{Massless spin-2 fields via lower spins}

\author{\"{O}zg\"{u}r A\c{c}{\i}k $^{1}$}
\email{ozacik@science.ankara.edu.tr}
\author{\"{U}mit Ertem $^{2}$}
\email{umitertemm@gmail.com}
\address{$^{1}$ Department of Physics, Ankara University,\\ Faculty of Sciences, 06100, Tando\u gan-Ankara,
Turkey\\
$^{2}$ Astronomer, Diyanet \.{I}\c{s}leri Ba\c{s}kanl{\i}\u{g}{\i}, \"{U}niversiteler Mah.\\
 Dumlup{\i}nar Bul. No:147/H 06800 \c{C}ankaya, Ankara, Turkey}

\begin{abstract}

Solutions of massless spin-2 field equations are written in terms of massless spin-$\frac{3}{2}$ Rarita-Schwinger fields and twistor spinors. It is shown that the proposed massless spin-2 fields satisfy the tracelessness and divergencelessness conditions and are in the kernel of the Laplace-Beltrami operator. A spin lowering procedure for special cases and a symmetry operator for massless spin-2 fields are also obtained. Description of massless spin-2 fields in terms of lower spin fields are found.

\end{abstract}

\maketitle

\section{Introduction}

For different spins, solutions of massless field equations can be mapped onto each other by spin raising and lowering procedures in conformally flat backgrounds \cite{Penrose Rindler}. In general, this can be done for lower spins for which the field equations determine the massless fields as being in the kernel of some first-order differential operators. However, for the higher spins, the procedure does not work automatically since the solutions must satisfy some extra constraints other than being in the kernel of some first-order operator. For the massless spin-0, spin-$\frac{1}{2}$ and spin-1 fields, the spin raising operators transforming massless spin-$s$ fields to spin-$\left(s+\frac{1}{2}\right)$ fields and spin lowering operators transforming spin-$s$ fields to spin-$\left(s-\frac{1}{2}\right)$ fields are constructed by using twistor spinors \cite{Charlton, Benn Charlton Kress, Benn Kress}. Twistor spinors are special types of spinors which determine the conformal properties of the background. They also appear as supersymmetry generators in superconformal field theories and conformal hidden symmetries in various backgrounds \cite{Cassani Klare Martelli Tomasiello Zaffaroni, de Medeiros, Ertem1, Ertem2}. For massless spin-$\frac{3}{2}$ Rarita-Schwinger fields, there is an extra tracelesness condition and the construction of spin raising and lowering operators has been done in \cite{Acik Ertem1}. For higher spins, especially for massless spin-2 field or graviton, more extra conditions appear and the spin raising and lowering prodecures have to be considered separately.

In this paper, we consider massless spin-2 field equations describing graviton which contain tracelessness and divergencelessness conditions in addition to being in the kernel of the Laplace-Beltrami operator. By using a twistor spinor and a massless spin-$\frac{3}{2}$ Rarita-Schwinger field, we construct a massless spin-2 field via the spin raising procedure. We prove that the constructed massless spin-2 field satisfy the tracelessness and divergencelessness conditions and is in the kernel of the Laplace-Beltrami operator for Ricci-flat backgrounds which arise from the gauge invariance property of the Rarita-Schwinger fields. This extends the previous spin raising constructions to massless spin-2 fields. We also construct a massless spin-$\frac{3}{2}$ Rarita-Schwinger field from a twistor spinor and a massless spin-2 field via the spin lowering procedure for the special case of four dimensions. From the combination of spin raising and lowering procedures, we find a symmetry operator for massless spin-2 fields which takes a massless spin-2 field solution and give another solution. By using the spin raising and lowering constructions for lower spin fields, we also find the descriptions of massless spin-2 fields in terms of twistor spinors and lower spin fields that are spin-1 Maxwell, spin-$\frac{1}{2}$ Dirac and spin-0 scalar fields.

The paper is organized as follows. In Section II, we propose a massless spin-2 field in terms of a twistor spinor and a massless spin-$\frac{3}{2}$ Rarita-Schwinger field and prove that it satisfy the tracelessness and divergencelessness conditions and is in the kernel of the Laplace-Beltrami operator. We also construct spin lowering operator from a massless spin-2 field to a massless spin-$\frac{3}{2}$ Rarita-Schwinger field and find a symmetry operator for massless spin-2 fields. Section III contains the construction of massless spin-2 fileds in terms of lower spin fields. Section IV concludes the paper. In two appendices, we give the definition of spinor inner products and the relation between Clifford calculus and gamma matrix notations.

\section{Massless spin-2 fields via Rarita-Schwinger fields}

We consider a symmetric tensor field $h$ which describes a massless spin-2 particle. In the coframe basis $e^a$ of the background manifold, it can be written as
\begin{eqnarray}
h&=&h_a\otimes e^a\nonumber\\
&=&h_{ab}e^a\otimes e^b.
\end{eqnarray}
Symmetric tensor $h_{ab}$ can be written in terms of the differential 1-form $h_a$ as
\begin{equation}
h_{ab}=i_{X_b}h_a
\end{equation}
with the property $i_{X_b}h_a=i_{X_a}h_b$ where $i_{X_a}$ corresponds to the contraction operator with respect to the frame basis $X_a$. For example, the force carrier particle of gravitation that is graviton can be described as a massless spin-2 field and can be written as a perturbation around the fixed background metric $\eta_{ab}$. So, the total metric becomes
\begin{equation}
g_{ab}=\eta_{ab}+h_{ab}.
\end{equation}
Symmetric tensor field $h$ must be traceless and divergenceless to describe a massless spin-2 particle. Moreover, it must be in the kernel of the Laplace-Beltrami operator. Then, the field equations of massless spin-2 fields can be written as follows
\begin{eqnarray}
\Delta h_a&=&0\\
\delta h_a&=&0\\
i_{X^a}h_a&=&0
\end{eqnarray}
where $\delta$ denotes the coderivative operator and $\Delta$ is the Laplace-Beltrami operator which can be written as
\begin{equation}
\Delta=-d\delta-\delta d
\end{equation}
in terms of exterior derivative operator $d$ and coderivative operator $\delta$.

We will construct massless spin-2 fields satisfying the field equations (4)-(6) in terms of lower spin fields by the method of spin raising. On a manifold admitting a twistor spinor, lower spin fields and twistor spinors can be combined to form a higher spin field by a spin raising operator. For the massless spin-2 case, the spin raising operator will be constructed from a twistor spinor and a spin-$\frac{3}{2}$ massless Rarita-Schwinger field. Twistor spinors are special types of spinors which are in the kernel of the following twistor (Penrose) operator defined on an $n$-dimensional spin manifold with respect to a vector field $X$ and its metric dual 1-form $\widetilde{X}$
\begin{equation}
\nabla_X-\frac{1}{n}\widetilde{X}.\displaystyle{\not}D.
\end{equation}
Here $\displaystyle{\not}D$ is the Dirac operator defined by
\begin{equation}
\displaystyle{\not}D=e^a.\nabla_{X_a}
\end{equation}
where $.$ denotes the Clifford product. So, a twistor spinor $u$ is a solution of the following equation
\begin{equation}
\nabla_{X_a}u=\frac{1}{n}e_a.\displaystyle{\not}Du.
\end{equation}
Taking second covariant derivative of the twistor equation will give the following integrability conditions
\begin{eqnarray}
\nabla_{X_a}\displaystyle{\not}Du&=&\frac{n}{2}K_a.u\\
\displaystyle{\not}D^2u&=&-\frac{n}{4(n-1)}{\cal{R}}u\\
W_{ab}.u&=&0.
\end{eqnarray}
Here, $K_a$ is the Schouten 1-form defined by
\begin{equation}
K_a=\frac{1}{n-2}\left(\frac{\cal{R}}{2(n-1)}e_a-P_a\right)
\end{equation}
in terms of Ricci 1-form $P_a$ and curvature scalar ${\mathcal{R}}$. $W_{ab}$ is Weyl (conformal) 2-form defined for $n>2$ by
\begin{equation}
W_{ab}=R_{ab}-\frac{1}{n-2}\left(P_a\wedge e_b-P_b\wedge e_a\right)+\frac{1}{(n-1)(n-2)}{\cal{R}}e_a\wedge e_b
\end{equation}
where $R_{ab}$ are curvature 2-forms. Note that, for the special case of Ricci-flat manifolds which satisfy the conditions $P_a=0={\mathcal{R}}$, the right hand sides of (11) and (12) vanish.

Spin-$\frac{3}{2}$ fields are described by spinor-valued 1-forms which are written as a tensor product of a spinor and a 1-form. We consider the following spinor-valued 1-form
\begin{equation}
\Psi=\psi_a\otimes e^a
\end{equation}
where $\psi_a$ is a spinor and $e^a$ are the coframe basis 1-forms. For $\Psi$ to be a massless Rarita-Schwinger field, it must satisfy the following field equations
\begin{eqnarray}
{\mathbb{\displaystyle{\not}D}}\Psi&=&0\\
e^a.\psi_a&=&0
\end{eqnarray}
where ${\mathbb{\displaystyle{\not}D}}$ is the Rarita-Schwinger operator acting on spinor-valued 1-forms and defined by
\begin{equation}
{\mathbb{\displaystyle{\not}D}}=e^a.\nabla_{X_a}
\end{equation}
and (18) corresponds to the tracelessness condition. Moreover, $\psi_a$ satisfies the following Lorentz-type and massless Dirac conditions
\begin{eqnarray}
\nabla_{X^a}\psi_a&=&0\nonumber\\
\displaystyle{\not}D\psi_a&=&0
\end{eqnarray}
and the gauge invariance of the massless Rarita-Schwinger field equations implies the Ricci-flatness of the background manifold \cite{acik Ertem1}. Namely, to have a massless spin-$\frac{3}{2}$ Rarita-Schwinger field satisfying (17) and (18), we must have a Ricci-flat background, $P_a=0$.

Now, we propose a 1-form constructed out of a twistor spinor $u$ and a massless spin-$\frac{3}{2}$ Rarita-Schwinger field $\Psi=\psi_a\otimes e^a$ by the spin raising procedure and we claim that this 1-form will correspond to a massless spin-2 field satisfying (4)-(6). On an $n$-dimensional Ricci-flat background, the proposed 1-form is the following
\begin{equation}
h_a=(e^b.u,e_c.\nabla_{X_b}\psi_a)e^c+\frac{n-2}{n}(\displaystyle{\not}Du,e_c.\psi_a)e^c+(\psi_a,e_c.\displaystyle{\not}Du)e^c
\end{equation}
where $(\,,\,)$ corresponds to the inner product for spinor fields defined in Appendix A.

\subsection{Proof of tracelessness}

First, we will prove the tracelessness condition given in (6). Contraction operator can be written in terms of Clifford products since we have the following identities for the Clifford product in terms of wedge product and contraction operator for the coframe basis $e^a$ and an arbitrary $p$-form $\alpha$
\begin{eqnarray}
e^a.\alpha&=&e^a\wedge\alpha+i_{X^a}\alpha\\
\alpha.e^a&=&e^a\wedge\eta\alpha-i_{X^a}\eta\alpha
\end{eqnarray}
where $\eta$ acts on a $p$-form $\alpha$ as $\eta\alpha=(-1)^p\alpha$. So, for the 1-form $h_a$ given in (21), we have
\begin{eqnarray}
e^a.h_a&=&e^a\wedge h_a+i_{X^a}h_a\\
h_a.e^a&=&-e^a\wedge h_a+i_{X^a}h_a
\end{eqnarray}
and by adding them to each other we obtain
\begin{equation}
i_{X^a}h_a=\frac{1}{2}(e^a.h_a+h_a.e^a).
\end{equation}
From the definiton of $h_a$ in (21), we can write
\begin{equation}
e^a.h_a=(e^b.u,e_c.\nabla_{X_b}\psi_a)e^a.e^c+\frac{n-2}{n}(\displaystyle{\not}Du,e_c.\psi_a)e^a.e^c+(\psi_a,e_c.\displaystyle{\not}Du)e^a.e^c
\end{equation}
and
\begin{equation}
h_a.e^a=(e^b.u,e_c.\nabla_{X_b}\psi_a)e^c.e^a+\frac{n-2}{n}(\displaystyle{\not}Du,e_c.\psi_a)e^c.e^a+(\psi_a,e_c.\displaystyle{\not}Du)e^c.e^a.
\end{equation}
By using the Clifford algebra identity
\begin{equation}
e^a.e^c+e^c.e^a=2g^{ac}
\end{equation}
in terms of the inverse metric $g^{ac}$, half of the sum of (27) and (28) gives
\begin{equation}
\frac{1}{2}(e^a.h_a+h_a.e^a)=(e^b.u,e^a.\nabla_{X_b}\psi_a)+\frac{n-2}{n}(\displaystyle{\not}Du,e^a.\psi_a)+(\psi_a,e^a.\displaystyle{\not}Du).
\end{equation}
Since $\psi_a$ is the spinor part of a massless Rarita-Schwinger field, it satisfies $e^a.\psi_a=0$ from (18) and the second term of (30) on the right hand side vanishes. Moreover, for any two spinors $\psi$ and $\phi$, we have the following spinor inner product property
\begin{equation}
(\psi,e^a.\phi)=((e^a)^{\mathcal{J}}.\psi,\phi)
\end{equation}
for the spinor inner product involution ${\mathcal{J}}$ whose properties are given in Appendix A. Depending on the chosen involution ${\mathcal{J}}$, we have $(e^a)^{\mathcal{J}}=\pm e^a$. Then, (30) transforms into
\begin{equation}
\frac{1}{2}(e^a.h_a+h_a.e^a)=\pm(e^a.e^b.u,\nabla_{X_b}\psi_a)\pm(e^a.\psi_a,\displaystyle{\not}Du)
\end{equation}
and from (18), the second term on the right hand side again vanishes. From the Clifford algebra identity (29), we can write $\pm e^a.e^b=\mp e^b.e^a\pm 2g^{ab}$ and obtain
\begin{eqnarray}
i_{X^a}h_a&=&\frac{1}{2}(e^a.h_a+h_a.e^a)\nonumber\\
&=&\mp(e^b.e^a.u,\nabla_{X_b}\psi)\pm 2(u,\nabla_{X^a}\psi_a)\nonumber\\
&=&-(e^a.u,e^b.\nabla_{X_b}\psi_a)\nonumber\\
&=&-(e^a.u,\displaystyle{\not}D\psi_a)\nonumber\\
&=&0
\end{eqnarray}
where we have used (31) and (20) in the third line, (9) in the fourth line and (20) in the last line. This proves the tracelessness condition for (21).

\subsection{Proof of divergencelessness}

To prove the divergencelessness of $h_a$ given in (21), we will calculate the coderivative of $h_a$. Coderivative operator $\delta$ can be written in terms of covariant derivative and contraction operator as
\begin{equation}
\delta=-i_{X^a}\nabla_{X_a}
\end{equation}
for a torsion-free connection. So, we have
\begin{eqnarray}
\delta h_a&=&-i_{X^b}\nabla_{X_b}\left[(e^d.u,e_c.\nabla_{X_d}\psi_a)e^c+\frac{n-2}{n}(\displaystyle{\not}Du,e_c.\psi_a)e^c+(\psi_a,e_c.\displaystyle{\not}Du)e^c\right]\nonumber\\
&=&-i_{X^b}\bigg[(e^d.\nabla_{X_b}u,e_c.\nabla_{X_d}\psi_a)e^c+(e^d.u,e_c.\nabla_{X_b}\nabla_{X_d}\psi_a)e^c+\frac{n-2}{n}(\nabla_{X_b}\displaystyle{\not}Du,e_c.\psi_a)e^c\nonumber\\
&&+\frac{n-2}{n}(\displaystyle{\not}Du,e_c.\nabla_{X_b}\psi_a)e^c+(\nabla_{X_b}\psi_a,e_c.\displaystyle{\not}Du)e^c+(\psi_a,e_c.\nabla_{X_b}\displaystyle{\not}Du)e^c\bigg]
\end{eqnarray}
where we have used the compatibility of the spinor inner product with the covariant derivative and normal coordinates satisfying $\nabla_{X_a}e^b=0$. From the identity $i_{X_b}e^c=\delta_b^c$ in terms of Kronecker delta, we can write
\begin{eqnarray}
\delta h_a&=&-\bigg[(e^d.\nabla_{X_b}u,e^b.\nabla_{X_d}\psi_a)+(e^d.u,e^b.\nabla_{X_b}\nabla_{X_d}\psi_a)+\frac{n-2}{n}(\nabla_{X_b}\displaystyle{\not}Du,e^b.\psi_a)\nonumber\\
&&+\frac{n-2}{n}(\displaystyle{\not}Du,e^b.\nabla_{X_b}\psi_a)+(\nabla_{X_b}\psi_a,e^b.\displaystyle{\not}Du)+(\psi_a,e^b.\nabla_{X_b}\displaystyle{\not}Du)\bigg]
\end{eqnarray}
and by using the definition of Dirac operator (9), the twistor equation (10) and the inner product property (31), we obtain
\begin{eqnarray}
\delta h_a&=&-\frac{1}{n}(e^d.e_b.\displaystyle{\not}Du,e^b.\nabla_{X_d}\psi_a)\mp(e^b.e^d.u,\nabla_{X_b}\nabla_{X_d}\psi_a)\mp\frac{n-2}{n}(\displaystyle{\not}D^2u,\psi_a)\nonumber\\
&&-\frac{n-2}{n}(\displaystyle{\not}Du,\displaystyle{\not}D\psi_a)\mp(\displaystyle{\not}D\psi_a,\displaystyle{\not}Du)-(\psi_a,\displaystyle{\not}D^2u).
\end{eqnarray}
From the condition (20) for Rarita-Schwinger fields, we have
\begin{equation}
\delta h_a=-\frac{1}{n}(e^d.e_b.\displaystyle{\not}Du,e^b.\nabla_{X_d}\psi_a)\mp(e^b.e^d.u,\nabla_{X_b}\nabla_{X_d}\psi_a)\mp\frac{n-2}{n}(\displaystyle{\not}D^2u,\psi_a)-(\psi_a,\displaystyle{\not}D^2u).
\end{equation}
The first term on the right hand side of (38) vanishes as can be seen in the following way
\begin{eqnarray}
(e^d.e_b.\displaystyle{\not}Du,e^b.\nabla_{X_d}\psi_a)&=&\pm(e^b.e^d.e_b.\displaystyle{\not}Du,\nabla_{X_d}\psi_a)\nonumber\\
&=&\mp(n-2)(e^d.\displaystyle{\not}Du,\nabla_{X_d}\psi_a)\nonumber\\
&=&-(n-2)(\displaystyle{\not}Du,\displaystyle{\not}D\psi_a)\nonumber\\
&=&0
\end{eqnarray}
where we have used the identity $e^b.\alpha.e_b=(-1)^p(n-2p)\alpha$ for a $p$-form $\alpha$, the property (31) and the condition (20). Then, we have
\begin{equation}
\delta h_a=\mp(e^b.e^d.u,\nabla_{X_b}\nabla_{X_d}\psi_a)\mp\frac{n-2}{n}(\displaystyle{\not}D^2u,\psi_a)-(\psi_a,\displaystyle{\not}D^2u).
\end{equation}
From (22), we can write Clifford product of coframe bases in terms of wedge product as $e^b.e^d=e^b\wedge e^d+g^{bd}$ and (40) transforms into
\begin{eqnarray}
\delta h_a&=&\mp((e^b\wedge e^d).u,\nabla_{X_b}\nabla_{X_d}\psi_a)\mp(u,\nabla_{X_b}\nabla_{X^b}\psi_a)\mp\frac{n-2}{n}(\displaystyle{\not}D^2u,\psi_a)-(\psi_a,\displaystyle{\not}D^2u)\nonumber\\
&=&\mp\frac{1}{2}((e^b\wedge e^d).u,(\nabla_{X_b}\nabla_{X_d}-\nabla_{X_d}\nabla_{X_b})\psi_a)\mp(u,\nabla^2\psi_a)\mp\frac{n-2}{n}(\displaystyle{\not}D^2u,\psi_a)-(\psi_a,\displaystyle{\not}D^2u)\nonumber\\
&=&\mp\frac{1}{2}((e^b\wedge e^d).u,R(X_b,X_d)\psi_a)\mp(u,\nabla^2\psi_a)\mp\frac{n-2}{n}(\displaystyle{\not}D^2u,\psi_a)-(\psi_a,\displaystyle{\not}D^2u)
\end{eqnarray}
where we take the antisymmetric part of the first term since it is antisymmetric in indices $b$ and $d$ and we use the definitions of the curvature operator $R(X_b,X_d)=[\nabla_{X_b},\nabla_{X_d}]$ and Laplacian $\nabla^2=\nabla_{X_b}\nabla_{X^b}$ in normal coordinates. The action of the curvature operator on a spinor can be written in terms of curvature 2-forms as \cite{Benn Tucker}
\begin{equation}
R(X_b,X_d)\psi_a=\frac{1}{2}R_{bd}.\psi_a
\end{equation}
and the Laplacian acting on a spinor can be written in terms of the square of the Dirac operator and scalar curvature because of the following Schr\"odinger-Lichnerowicz-Weitzenb\"ock identity
\begin{equation}
\displaystyle{\not}D^2\psi_a=\nabla^2\psi_a-\frac{1}{4}{\mathcal{R}}\psi_a.
\end{equation}
So, we can write the coderivative of $h_a$ as
\begin{equation}
\delta h_a=\mp\frac{1}{4}((e^b\wedge e^d).u,R_{bd}.\psi_a)\mp\frac{1}{4}({\mathcal{R}}u,\psi_a)\mp\frac{n-2}{n}(\displaystyle{\not}D^2u,\psi_a)-(\psi_a,\displaystyle{\not}D^2u)
\end{equation}
where we have used (20). We can write the curvature 2-forms as $R_{bd}=\frac{1}{2}(e^p\wedge e^q)i_{X_q}i_{X_p}R_{bd}$ and because of the pairwise symmetry of the curvature 2-forms $i_{X_c}i_{X_d}R_{ab}=i_{X_a}i_{X_b}R_{cd}$, the first term on the right hand side can be written as
\[
((e^b\wedge e^d).u,R_{bd}.\psi_a)=(R_{bd}.u,(e^b\wedge e^d).\psi_a).
\]
From the manifestation of curvature 2-forms in terms of Weyl 2-forms and Schouten 1-forms
\begin{equation}
R_{ab}=W_{ab}+e_b.K_a-e_a.K_b
\end{equation}
and by using the integrability conditions (11) and (13) of twistor spinors, the action of curvature 2-forms on a twistor spinor can be written as \cite{Ertem3}
\begin{equation}
R_{bd}.u=\frac{2}{n}\left(e_d.\nabla_{X_b}\displaystyle{\not}Du-e_b.\nabla_{X_d}\displaystyle{\not}Du\right).
\end{equation}
Moreover, by using the integrability condition (12) for twistor spinors and the property (31) for spinor inner products, we obtain for (44)
\begin{eqnarray}
\delta h_a&=&\mp\frac{1}{2n}(e_d.\nabla_{X_b}\displaystyle{\not}Du-e_b.\nabla_{X_d}\displaystyle{\not}Du,(e^b\wedge e^d).\psi_a)\mp\frac{1}{4}({\mathcal{R}}u,\psi_a)\pm\frac{n-2}{4(n-1)}({\mathcal{R}}u,\psi_a)\mp(e^b.\psi_a,\nabla_{X_b}\displaystyle{\not}Du)\nonumber\\
&=&\mp\frac{1}{2n}(e_d.\nabla_{X_b}\displaystyle{\not}Du-e_b.\nabla_{X_d}\displaystyle{\not}Du,(e^b\wedge e^d).\psi_a)\mp\frac{1}{4(n-1)}({\mathcal{R}}u,\psi_a)\mp(e^b.\psi_a,\nabla_{X_b}\displaystyle{\not}Du).
\end{eqnarray}
By writing the wedge product of coframe bases in terms of Clifford product $e^b\wedge e^d=e^b.e^d-g^{bd}$, the first term on the right hand side of (47) transforms into
\begin{eqnarray}
\mp\frac{1}{2n}(e_d.\nabla_{X_b}\displaystyle{\not}Du-e_b.\nabla_{X_d}\displaystyle{\not}Du,(e^b\wedge e^d).\psi_a)&=&\mp\frac{1}{2n}(e_d.\nabla_{X_b}\displaystyle{\not}Du-e_b.\nabla_{X_d}\displaystyle{\not}Du,(e^b.e^d).\psi_a)\nonumber\\
&=&\mp\frac{1}{2n}(e_d.\nabla_{X_b}\displaystyle{\not}Du,(e^b.e^d-e^d.e^b).\psi_a)\nonumber\\
&=&\pm\frac{1}{n}(e_d.\nabla_{X_b}\displaystyle{\not}Du,e^d.e^b.\psi_a)\mp\frac{1}{n}(\displaystyle{\not}D^2u,\psi_a)\nonumber\\
&=&\pm\frac{1}{n}(e_d.\nabla_{X_b}\displaystyle{\not}Du,e^d.e^b.\psi_a)\pm\frac{1}{4(n-1)}({\mathcal{R}}u,\psi_a)
\end{eqnarray}
where we renamed the indices in the second line, used the identity $e^b.e^d=-e^d.e^b+2g^{bd}$ in the third line and the integrability condition (12) in the last line. Then, we obtain
\begin{eqnarray}
\delta h_a&=&\pm\frac{1}{n}(e_d.\nabla_{X_b}\displaystyle{\not}Du,e^d.e^b.\psi_a)\pm\frac{1}{4(n-1)}({\mathcal{R}}u,\psi_a)\mp\frac{1}{4(n-1)}({\mathcal{R}}u,\psi_a)\mp(e^b.\psi_a,\nabla_{X_b}\displaystyle{\not}Du)\nonumber\\
&=&\pm\frac{1}{n}(e_d.\nabla_{X_b}\displaystyle{\not}Du,e^d.e^b.\psi_a)\mp(e^b.\psi_a,\nabla_{X_b}\displaystyle{\not}Du)\nonumber\\
&=&(\nabla_{X_b}\displaystyle{\not}Du,e^b.\psi_a)\mp(e^b.\psi_a,\nabla_{X_b}\displaystyle{\not}Du)
\end{eqnarray}
where we have used the inner product property (31) and the identity $e^d.e_d=n$. From the definitions given in Appendix A, if we choose the induced involution $j$ of spinor inner product as identity, we can reverse the order of terms in the inner product depending on we have a symmetric or skewsymmetric inner product as
\begin{equation}
(\nabla_{X_b}\displaystyle{\not}Du,e^b.\psi_a)=\epsilon(e^b.\psi_a,\nabla_{X_b}\displaystyle{\not}Du)
\end{equation}
where $\epsilon=+1$ for symmetric inner product and $\epsilon=-1$ for skewsymmetric inner product. So, we have
\begin{equation}
\delta h_a=\epsilon(e^b.\psi_a,\nabla_{X_b}\displaystyle{\not}Du)\mp(e^b.\psi_a,\nabla_{X_b}\displaystyle{\not}Du).
\end{equation}
The terms on the right hand side of (51) are same up to a sign factor. The factor of the first term comes from the symmetry or skewsymmetry of the inner product and the factor of the second term comes from the reverse sign of $(e^a)^{\mathcal{J}}$. Now, we can analyze the cases for which the divergence of $h_a$ vanishes. From the classification of spinor inner products in various dimensions given in \cite{Ertem Sutemen Acik Catalkaya}, there are three possible cases that this can happen. For Lorentzian 4-manifolds, we can choose the spinor inner product as $\mathbb{C}$-skew with $\xi\eta$ involution. In that case, we have $\epsilon=-1$ and $(e^a)^{\xi\eta}=-e^a$ and the terms on the right hand side of (51) cancel each other. For Lorentzian 10-manifolds, we can choose the spinor inner product as $\mathbb{R}$-sym$\oplus\mathbb{R}$-sym with $\xi$ involution. In that case, we have $\epsilon=+1$ and $(e^a)^{\xi}=e^a$ and the terms on the right hand side of (51) cancel each other. For Lorentzian 11-manifolds, we can choose the spinor inner product as $\mathbb{R}$-sym with $\xi$ involution or $\mathbb{R}$-skew with $\xi\eta$ involution. In that case, we have $\epsilon=+1$ and $(e^a)^{\xi\eta}=e^a$ or $\epsilon=-1$ and $(e^a)^{\xi\eta}=-e^a$ and for both cases the terms on the right hand side of (51) cancel each other. So, the divergencelessness property is in consistence with 4, 10 and 11-dimensional supergravity theories. This analysis is relevant for general background manifolds. However, remember that to have a Rarita-Schwinger solution, we need Ricci-flat backgrounds and from the integrability condition (11), both terms on the right hand side of (51) automatically vanishes in that case. So, for all Ricci-flat manifolds, we have
\begin{equation}
\delta h_a=0
\end{equation}
and this proves the divergencelessness property of $h_a$ given in (21).

\subsection{Proof of being in the kernel of Laplace-Beltrami operator}

Now, we can consider (4) which corresponds to $h_a$ being in the kernel of the Laplace-Beltrami operator. Since the coderivative of $h_a$ vanishes, the action of the Laplace-Beltrami operator $\Delta$ defined in (7) to $h_a$ reduces to
\begin{equation}
\Delta h_a=-\delta dh_a.
\end{equation}
Let us first calculate the exterior derivative of $h_a$ by using the following expression of $d$ for a torsion-free connection
\begin{equation}
d=e^b\wedge \nabla_{X_b}.
\end{equation}
The covariant derivative of (21), which is expressed in (35), gives
\begin{eqnarray}
\nabla_{X_b}h_a&=&(e^d.\nabla_{X_b}u,e_c.\nabla_{X_d}\psi_a)e^c+(e^d.u,e_c.\nabla_{X_b}\nabla_{X_d}\psi_a)e^c+\frac{n-2}{n}(\nabla_{X_b}\displaystyle{\not}Du,e_c.\psi_a)e^c\nonumber\\
&&+\frac{n-2}{n}(\displaystyle{\not}Du,e_c.\nabla_{X_b}\psi_a)e^c+(\nabla_{X_b}\psi_a,e_c.\displaystyle{\not}Du)e^c+(\psi_a,e_c.\nabla_{X_b}\displaystyle{\not}Du)e^c\nonumber\\
&=&\frac{1}{n}(e^d.e_b.\displaystyle{\not}Du,e_c.\nabla_{X_d}\psi_a)e^c+(e^d.u,e_c.\nabla_{X_b}\nabla_{X_d}\psi_a)e^c\nonumber\\
&&+(\nabla_{X_b}\psi_a,e_c.\displaystyle{\not}Du)e^c\pm\frac{n-2}{n}(e_c.\displaystyle{\not}Du, \nabla_{X_b}\psi_a)e^c
\end{eqnarray}
where we have used the twistor equation (10), the integrability condition (11) for Ricci-flat backgrounds and the properties of the spinor inner product. From (54), we can write the exterior derivative of $h_a$ as
\begin{eqnarray}
dh_a&=&\frac{1}{n}(e^d.e_b.\displaystyle{\not}Du,e_c.\nabla_{X_d}\psi_a)e^b\wedge e^c+(e^d.u,e_c.\nabla_{X_b}\nabla_{X_d}\psi_a)e^b\wedge e^c\nonumber\\
&&+(\nabla_{X_b}\psi_a,e_c.\displaystyle{\not}Du)e^b\wedge e^c\pm\frac{n-2}{n}(e_c.\displaystyle{\not}Du, \nabla_{X_b}\psi_a)e^b\wedge e^c.
\end{eqnarray}
The first term on the right hand side of (56) can be expressed as
\begin{eqnarray}
(e^d.e_b.\displaystyle{\not}Du,e_c.\nabla_{X_d}\psi_a)e^b\wedge e^c&=&\pm(e_b.\displaystyle{\not}Du, e^d.e_c.\nabla_{X_d}\psi_a)e^b\wedge e^c\nonumber\\
&=&\mp(e_b.\displaystyle{\not}Du,e_c.\displaystyle{\not}D\psi_a)e^b\wedge e^c\pm 2(e_b.\displaystyle{\not}Du,\nabla_{X_c}\psi_a)e^b\wedge e^c\nonumber\\
&=&\pm 2(e_b.\displaystyle{\not}Du,\nabla_{X_c}\psi_a)e^b\wedge e^c
\end{eqnarray}
where we have used the identity $e^d.e_c=-e_c.e^d+2g^d_c$ with (9) and (20). Similarly, the second term on the right hand side of (56) can be written as
\begin{eqnarray}
(e^d.u,e_c.\nabla_{X_b}\nabla_{X_d}\psi_a)e^b\wedge e^c&=&\pm(e_c.e^d.u,\nabla_{X_b}\nabla_{X_d}\psi_a)e^b\wedge e^c\nonumber\\
&=&\pm((e_c\wedge e^d).u,\nabla_{X_b}\nabla_{X_d}\psi_a)e^b\wedge e^c\pm(u,\nabla_{X_b}\nabla_{X_c}\psi_a)e^b\wedge e^c\nonumber\\
&=&\pm((e_c\wedge e^d).u,\nabla_{X_b}\nabla_{X_d}\psi_a)e^b\wedge e^c\mp\frac{1}{4}(R_{bc}.u,\psi_a)e^b\wedge e^c\nonumber\\
&=&\pm((e_c\wedge e^d).u,\nabla_{X_b}\nabla_{X_d}\psi_a)e^b\wedge e^c
\end{eqnarray}
where we have used $e_c.e^d=e_c\wedge e^d+i_{X_c}e^d$, definition of the curvature operator, its action on spinors (42) and the equality $(u,R_{bc}.\psi_a)=-(R_{bc}.u,\psi_a)$ with the identity (46) and integrability condition (11) in Ricci-flat backgrounds. By rearranging the indices in (56), we obtain
\begin{eqnarray}
dh_a&=&\pm\frac{n-4}{n}(e_c.\displaystyle{\not}Du, \nabla_{X_b}\psi_a)e^b\wedge e^c+(\nabla_{X_b}\psi_a,e_c.\displaystyle{\not}Du)e^b\wedge e^c\nonumber\\
&&\pm((e_c\wedge e^d).u,\nabla_{X_b}\nabla_{X_d}\psi_a)e^b\wedge e^c.
\end{eqnarray}
By using the identities $e_c\wedge e^d=e_c.e^d-i_{X_c}e^d$ and $e_c.e^d=-e^d.e_c+2{g_c}^d$, the last term on the right hand side of (59) can be written as
\begin{eqnarray}
((e_c\wedge e^d).u,\nabla_{X_b}\nabla_{X_d}\psi_a)e^b\wedge e^c&=&-(e^d.e_c.u,\nabla_{X_b}\nabla_{X_d}\psi_a)e^b\wedge e^c\nonumber\\
&=&-(e^d.e_c.u,R(X_b,X_d)\psi_a)e^b\wedge e^c-(e^d.e_c.u,\nabla_{X_d}\nabla_{X_b}\psi_a)e^b\wedge e^c\nonumber\\
&=&-\frac{1}{2}(e^d.e_c.u,R_{bd}.\psi_a)e^b\wedge e^c\mp(e_c.u,\displaystyle{\not}D\nabla_{X_b}\psi_a)e^b\wedge e^c\nonumber\\
&=&\pm\frac{1}{2}(e_c.u,P_b.\psi_a)e^b\wedge e^c\mp(e_c.u,\nabla_{X_b}\displaystyle{\not}D\psi_a)e^b\wedge e^c\nonumber\\
&=&0
\end{eqnarray}
where we have used the definition of the curvature operator in the second line, the action of it on spinors (42) and the definiton of the Dirac operator (9) in the third line, the identity $e^d.R_{bd}=-P_b$ and the commutativity of $\nabla_{X_b}$ and $\displaystyle{\not}D$ in Ricci-flat backgrounds in the fourth line and the Ricci-flatness and (20) in the last line. So, the exterior derivative of $h_a$ is found as
\begin{equation}
dh_a=\pm\frac{n-4}{n}(e_c.\displaystyle{\not}Du, \nabla_{X_b}\psi_a)e^b\wedge e^c+(\nabla_{X_b}\psi_a,e_c.\displaystyle{\not}Du)e^b\wedge e^c.
\end{equation}
Now, we can calculate the action of the Laplace-Beltrami operator on $h_a$ from (53). By using the definition of coderivative in (34), we have
\begin{eqnarray}
-\delta dh_a&=&i_{X^d}\nabla_{X_d}dh_a\nonumber\\
&=&\pm\frac{n-4}{n}i_{X^d}\bigg((e_c.\nabla_{X_d}\displaystyle{\not}Du, \nabla_{X_b}\psi_a)e^b\wedge e^c+(e_c.\displaystyle{\not}Du, \nabla_{X_d}\nabla_{X_b}\psi_a)e^b\wedge e^c\bigg)\nonumber\\
&&+i_{X^d}\bigg((\nabla_{X_d}\nabla_{X_b}\psi_a,e_c.\displaystyle{\not}Du)e^b\wedge e^c+(\nabla_{X_b}\psi_a,e_c.\nabla_{X_d}\displaystyle{\not}Du)e^b\wedge e^c\bigg)\nonumber\\
&=&\pm\frac{n-4}{n}i_{X^d}\bigg((e_c.\displaystyle{\not}Du, \nabla_{X_d}\nabla_{X_b}\psi_a)e^b\wedge e^c\bigg)+i_{X^d}\bigg((\nabla_{X_d}\nabla_{X_b}\psi_a,e_c.\displaystyle{\not}Du)e^b\wedge e^c\bigg)
\end{eqnarray}
where we have used normal coordinates and the integrability condition (11) in Ricci-flat case. Then, we have
\begin{eqnarray}
-\delta dh_a&=&\pm\frac{n-4}{n}\bigg((e_c.\displaystyle{\not}Du, \nabla_{X_d}\nabla_{X^d}\psi_a)e^c-(e^d.\displaystyle{\not}Du, \nabla_{X_d}\nabla_{X_b}\psi_a)e^b\bigg)\nonumber\\
&&+\bigg((\nabla_{X_d}\nabla_{X^d}\psi_a,e_c.\displaystyle{\not}Du)e^c-(\nabla_{X_d}\nabla_{X_b}\psi_a,e^d.\displaystyle{\not}Du)e^b\bigg)\nonumber\\
&=&\pm\frac{n-4}{n}\bigg((e_c.\displaystyle{\not}Du, \nabla^2\psi_a)e^c\mp(\displaystyle{\not}Du, \displaystyle{\not}D\nabla_{X_b}\psi_a)e^b\bigg)\nonumber\\
&&+\bigg((\nabla^2\psi_a,e_c.\displaystyle{\not}Du)e^c\mp(\displaystyle{\not}D\nabla_{X_b}\psi_a,\displaystyle{\not}Du)e^b\bigg).
\end{eqnarray}
From (43) and the commutativity of covariant derivative and Dirac operator on Ricci-flat backgrounds, we obtain
\begin{eqnarray}
-\delta dh_a&=&\pm\frac{n-4}{n}\bigg((e_c.\displaystyle{\not}Du, \displaystyle{\not}D^2\psi_a-\frac{1}{4}{\mathcal{R}}\psi_a)e^c\mp(\displaystyle{\not}Du, \nabla_{X_b}\displaystyle{\not}D\psi_a)e^b\bigg)\nonumber\\
&&+\bigg((\displaystyle{\not}D^2\psi_a-\frac{1}{4}{\mathcal{R}}\psi_a,e_c.\displaystyle{\not}Du)e^c\mp(\nabla_{X_b}\displaystyle{\not}D\psi_a,\displaystyle{\not}Du)e^b\bigg)\nonumber\\
&=&0
\end{eqnarray}
where we have used (20). Hence, this proves that the 1-form $h_a$ defined in (21) is in the kernel of the Laplace-Beltrami operator in Ricci-flat backgrounds. As a result, the 1-form $h_a$ constructed from a twistor spinor $u$ and a Rarita-Schwinger field $\psi_a$ is a solution of the massless spin-2 field equations in Ricci-flat backgrounds.

\subsection{Spin lowering}

In a reverse procedure to the above spin raising calculations, we can also search for a spin lowering process that transforms a massless spin-2 field to a massless spin-$\frac{3}{2}$ Rarita-Schwinger field. This can be done in the special case of 4-dimensions for a massless spin-2 field $h_a$ which additionally satisfies the following (anti)self-duality condition for the 2-form $dh_a$
\begin{equation}
*dh_a=\pm dh_a.
\end{equation}
In that case, by using a twistor spinor $u$, we can define a massless spin-$\frac{3}{2}$ field $\Psi=\psi_a\otimes e^a$ as
\begin{equation}
\psi_a=dh_a.u
\end{equation}
and it satisfies the field equations (17) and (18). This can be proven as follows. If we replace (66) in (18), we have
\begin{eqnarray}
e^a.\psi_a&=&e^a.dh_a.u\nonumber\\
&=&(e^a\wedge dh_a+i_{X^a}dh_a).u\nonumber\\
&=&(-d(e^a\wedge h_a)+de^a\wedge h_a+i_{X^a}dh_a).u
\end{eqnarray}
where we have used (22) and the antiderivative property of $d$. In normal coordinates we have $de^a=0$ and we can write $e^a\wedge h_a=h_{ab}e^a\wedge e^b=0$ since $h_{ab}$ is symmetric and $e^a\wedge e^b$ is antisymmetric. So, only the last term in (67) survives. Because of the square of the Hodge star operator is plus or minus identity $*^2=\pm1$, we can write
\begin{eqnarray}
i_{X^a}dh_a&=&\pm i_{X^a}**dh_a\nonumber\\
&=&\pm*(dh_a\wedge e^a)\nonumber\\
&=&\pm*(d(h_a\wedge e^a)-h_a\wedge de^a)\nonumber\\
&=&0
\end{eqnarray}
where we have used the identity $i_{X^a}*\alpha=*(\alpha\wedge e^a)$ for any form $\alpha$. This means that the right hand side of (67) vanishes and (66) satisfies (18). For the first field equation (17), it is enough to prove (20) for $\psi_a$ defined in (66). Then, we have
\begin{eqnarray}
\displaystyle{\not}D\psi_a&=&e^b.\nabla_{X_b}(dh_a.u)\nonumber\\
&=&e^b.\nabla_{X_b}dh_a.u+e^b.dh_a.\nabla_{X_b}u\nonumber\\
&=&\displaystyle{\not}ddh_a.u+\frac{1}{n}e^b.dh_a.e_b.\displaystyle{\not}Du
\end{eqnarray}
where we have used the definition of the Hodge-de Rham operator $\displaystyle{\not}d=e^b.\nabla_{X_b}$ on Clifford forms and the twistor equation (10). Moreover, from $\displaystyle{\not}d=d-\delta$ and the identity $e^a.\alpha.e_a=(-1)^p(n-2p)\alpha$ for a $p$-form $\alpha$, we have
\begin{eqnarray}
\displaystyle{\not}D\psi_a&=&-\delta dh_a.u+\frac{n-4}{n}dh_a.\displaystyle{\not}Du\nonumber\\
&=&\Delta h_a.u\nonumber\\
&=&0
\end{eqnarray}
from (53) and for $n=4$. As a result, (66) satisfies the massless Rarita-Schwinger field equations for a massless spin-2 field $h_a$ which satisfy the condition (65) in 4-dimensions and we have a spin lowering procedure for these special cases.

\subsection{Symmetry operators}

By using the above spin raising and lowering procedures, we can construct symmetry operators for massless spin-2 fields which transform a massless spin-2 field solution to another massless spin-2 field. Let us assume that we have a massless spin-2 field $h'_a$ in 4-dimensions which satisfy the condition (65). Then we can construct a massless Rarita-Schwinger field
\begin{equation}
\Psi=dh'_a.u_1\otimes e^a
\end{equation}
by using a twistor spinor $u_1$ as in (66). Now, we can replace this Rarita-Schwinger field in (21) to obtain another massless spin-2 field solution. So, we obtain
\begin{equation}
h_a=(e^b.u_2, e_c.\nabla_{X_b}(dh'_a.u_1))e^c+\frac{n-2}{n}(\displaystyle{\not}Du_2, e_c.dh'_a.u_1)e^c+(dh'_a.u_1, e_c.\displaystyle{\not}Du_2)e^c
\end{equation}
for another twistor spinor $u_2$. By using (10), we can write it as
\begin{eqnarray}
h_a&=&(e^b.u_2, e_c.\nabla_{X_b}dh'_a.u_1)e^c+\frac{1}{n}(e^b.u_2, e_c.dh' _a.e_b.\displaystyle{\not}Du_1)e^c\nonumber\\
&&+\frac{n-2}{n}(\displaystyle{\not}Du_2, e_c.dh'_a.u_1)e^c+(dh'_a.u_1, e_c.\displaystyle{\not}Du_2)e^c.
\end{eqnarray}
Moreover, we can write the first term on the right hand side of (73) as follows
\begin{eqnarray}
(e^b.u_2, e_c.\nabla_{X_b}dh'_a.u_1)&=&\pm(u_2, e^b.e_c.\nabla_{X_b}dh'_a.u_1)\nonumber\\
&=&\mp(u_2, e_c.e^b.\nabla_{X_b}dh'_a.u_1)\pm2(u_2, \nabla_{X_c}dh'_a.u_1)\nonumber\\
&=&\pm2(u_2, \nabla_{X_c}dh'_a.u_1)
\end{eqnarray}
where we have used the identity $e^b.e_c=-e_c.e^b+2g^b_c$ in the second line and $\displaystyle{\not}d=e^b.\nabla_{X_b}$ with $\displaystyle{\not}ddh'_a=-\delta dh'_a=0$ in the last line. The second term on the right hand side of (73) can also be written as
\begin{eqnarray}
(e^b.u_2, e_c.dh'_a.e_b.\displaystyle{\not}Du_1)&=&\pm(u_2, e^b.e_c.dh'_a.e_b.\displaystyle{\not}Du_1)\nonumber\\
&=&\mp(u_2, e_c.e^b.dh'_a.e_b.\displaystyle{\not}Du_1)\pm2(u_2, dh'_a.e_c.\displaystyle{\not}Du_1)\nonumber\\
&=&\pm2(u_2, dh'_a.e_c.\displaystyle{\not}Du_1)
\end{eqnarray}
where we have used the identity $e^b.e_c=-e_c.e^b+2g^b_c$ in the second line and $e^b.dh'_a.e_b=(n-4)dh'_a=0$ for $n=4$ in the last line.

As a result, we find a symmetry operator $L_{u_1u_2}$ between $h'_a$ and $h_a$ written in terms of two twistor spinors $u_1$ and $u_2$ in the following form
\begin{eqnarray}
h_a&=&L_{u_1u_2}h'_a\nonumber\\
&=&\pm2(u_2, \nabla_{X_b}dh'_a.u_1)e^b\pm\frac{1}{2}(u_2, dh'_a.e_b.\displaystyle{\not}Du_1)e^b\nonumber\\
&&+\frac{1}{2}(\displaystyle{\not}Du_2, e_b.dh'_a.u_1)e^b+(dh'_a.u_1, e_b.\displaystyle{\not}Du_2)e^b
\end{eqnarray}
which is relevant for $n=4$. Indeed, we can also write this symmetry operator in terms of only one twistor spinor $u$ if we take $u=u_1=u_2$ as follows
\begin{eqnarray}
h_a&=&L_uh'_a\nonumber\\
&=&\pm2(u, \nabla_{X_b}dh'_a.u)e^b\pm\frac{1}{2}(u, dh'_a.e_b.\displaystyle{\not}Du)e^b\nonumber\\
&&+\frac{1}{2}(\displaystyle{\not}Du, e_b.dh'_a.u)e^b+(dh'_a.u, e_b.\displaystyle{\not}Du)e^b\nonumber\\
&=&\pm2(u, \nabla_{X_b}dh'_a.u)e^b+\left(k\pm\frac{1}{2}\right)(u, dh'_a.e_b.\displaystyle{\not}Du)e^b\pm\frac{k}{2}(dh'_a.e_b.\displaystyle{\not}Du, u)e^b
\end{eqnarray}
where we have defined $(dh'_a)^{\mathcal{J}}=kdh'_a$ and used the previous convention $(e^b)^{\mathcal{J}}=\pm e^b$.

\section{Massless spin-2 fields via lower spin fields}

We can also write massless spin-2 fields in terms of lower than spin-$\frac{3}{2}$ fields. It is known that spin-$\frac{3}{2}$ massless Rarita-Schwinger fields can be written in terms of souce-free spin-1 field $F$ satisfying
\begin{equation}
\displaystyle{\not}dF=0
\end{equation}
and a twistor spinor $u$ in $n$-dimensions as \cite{Acik Ertem1}
\begin{equation}
\psi_a=\nabla_{X_a}F.u-\frac{1}{n}F.e_a.\displaystyle{\not}Du.
\end{equation}
If we replace this in (21) as the Rarita-Schwinger field, then we can write massless spin-2 field $h_a$ in terms of the souce-free spin-1 field $F$ and two twistor spinors $u_1$ and $u_2$. Hence, we obtain
\begin{eqnarray}
h_a&=&(e^b.u_1, e_c.\nabla_{X_b}(\nabla_{X_a}F.u_2))e^c-\frac{1}{n}(e^b.u_1, e_c.\nabla_{X_b}(F.e_a.\displaystyle{\not}Du_2))e^c+\frac{n-2}{n}(\displaystyle{\not}Du_1, e_c.\nabla_{X_a}F.u_2)e^c\nonumber\\
&&-\frac{n-2}{n^2}(\displaystyle{\not}Du_1, e_c.F.e_a.\displaystyle{\not}Du_2)e^c+(\nabla_{X_a}F.u_2, e_c.\displaystyle{\not}Du_1)e^c-\frac{1}{n}(F.e_a.\displaystyle{\not}Du_2, e_c.\displaystyle{\not}Du_1)e^c
\end{eqnarray}
By using the twistor equation (10), source-free spin-1 field equation (78), Ricci-flatness condition and the identities used along the paper, one can find
\begin{eqnarray}
h_a&=&\pm2(u_1, \nabla_{X_c}\nabla_{X_a}F.u_2)e^c+\frac{1}{n}(u_1, (\mp2\nabla_{X_c}F.e_a\pm2\nabla_{X_a}F.e_c\mp(n-4)e_c.\nabla_{X_a}F).\displaystyle{\not}Du_2)e^c\nonumber\\
&&+\frac{n-2}{n}(\displaystyle{\not}Du_1, e_c.\nabla_{X_a}F.u_2)e^c\pm(e_c.\nabla_{X_a}F.u_2, \displaystyle{\not}Du_1)e^c\nonumber\\
&&-\frac{n-2}{n^2}(\displaystyle{\not}Du_1, e_c.F.e_a.\displaystyle{\not}Du_2)e^c\mp\frac{1}{n}(e_c.F.e_a.\displaystyle{\not}Du_2, \displaystyle{\not}Du_1)e^c
\end{eqnarray}
or it can also be written in terms of just one twistor spinor by taking $u_1=u_2=u$.

We can also replace source-free Maxwell field $F$ in (81) with a massless spin-$\frac{1}{2}$ Dirac field $\psi$ and a twistor spinor $u$ \cite{Charlton}. We have the following expression for $F$ in terms of $\psi$ and $u$
\begin{equation}
F=(e^a.u, e_c.e_b.\nabla_{X_a}\psi)e^b\wedge e^c+\frac{n-2}{n}(\displaystyle{\not}Du,e_c.e_b.\psi)e^b\wedge e^c+(\psi, e_c.e_b.\displaystyle{\not}Du)e^b\wedge e^c,
\end{equation}
Moreover, massless spin-$\frac{1}{2}$ Dirac field can also be written in terms of a massless spin-0 scalar field $\phi$ and a twistor spinor $u$ as
\begin{equation}
\psi=d\phi.u+\frac{n-2}{n}\phi.\displaystyle{\not}Du
\end{equation}
and can be replaced in (82) to obtain an expression of massless spin-2 field $h_a$ in terms of massless spin-0 field $\phi$ and twistor spinors. In fact, we have the following transformations of massless fields via twistor spinors $u$
\begin{eqnarray}
\text{spin-}0\quad\phi\quad\quad&\rightarrow&\quad\quad\text{spin-}\frac{1}{2}\quad\psi=d\phi.u+\frac{n-2}{n}\phi.\displaystyle{\not}Du\nonumber\\
\text{spin-}\frac{1}{2}\quad\psi\quad\quad&\rightarrow&\quad\quad\text{spin-1}\quad F=(e^a.u, e_c.e_b.\nabla_{X_a}\psi)e^b\wedge e^c+\frac{n-2}{n}(\displaystyle{\not}Du,e_c.e_b.\psi)e^b\wedge e^c+(\psi, e_c.e_b.\displaystyle{\not}Du)e^b\wedge e^c\nonumber\\
\text{spin-}1\quad F\quad\quad&\rightarrow&\quad\quad\text{spin-}\frac{3}{2}\quad\psi_a=\nabla_{X_a}F.u-\frac{1}{n}F.e_a.\displaystyle{\not}Du\\
\text{spin-}\frac{3}{2}\quad\psi_a\quad\quad&\rightarrow&\quad\quad\text{spin-2}\quad h_a=(e^b.u,e_c.\nabla_{X_b}\psi_a)e^c+\frac{n-2}{n}(\displaystyle{\not}Du,e_c.\psi_a)e^c+(\psi_a,e_c.\displaystyle{\not}Du)e^c\nonumber
\end{eqnarray}

\section{Conclusion}

We show that the solutions of massless spin-2 field equations can be written as a 1-form in terms of massless spin-$\frac{3}{2}$ fields and twistor spinors by using spinor inner products. We prove that the proposed 1-form satisfies the tracelessness and divergencelessness conditions and is in the kernel of the Laplace-Beltrami operator. We also find a reverse procedure of spin lowering from massless spin-2 fields to massless spin-$\frac{3}{2}$ fields in four dimensions. The combination of these two constructions give a symmetry operator for massless spin-2 fields. As a result, we find a description of the spin-2 graviton field in terms of the massless Rarita-Schwinger fields and twistor spinors. This construction generalizes the previous constructions of spin raising and lowering procedures for lower spins.

One can search for generalizations of these spin raising and lowering procedures for more general backgrounds admitting gauged twistor spinors which are twistor spinors in the existence of the Spin$^c$ structures. Finding a general spin raising and lowering procedure encompassing all spins by using the special examples can also be investigated. Moreover, the search for a spin raising and lowering procedure for massive fields with different spins in terms of geometric Killing spinors is another topic of future research which can contain different restrictions on the backgrounds.

\begin{acknowledgments}

\"O. A. is suppoted by the Scientific and Technological Research Council of Turkey (T\"UB\.ITAK) Research Project No. 118F086.

\end{acknowledgments}

\begin{appendix}

\section{Inner products on spinor spaces}

Let us denote the spinor space by $S$. The elements of the dual spinor space $S^*$ maps the elements of $S$ to the corresponding division algebra $\mathbb{D}$ where $\mathbb{D}=\mathbb{R}, \mathbb{C}$ or $\mathbb{H}$ depending on the dimension and signature of the background. This defines an inner product $(\,,\,)$ on spinors. For two spinors $\psi, \phi$ and the dual spinor $\overline{\psi}$, we have
\begin{equation}
\overline{\psi}(\phi)=(\psi,\phi).
\end{equation}
There can be different types of spinor inner products on $S$ depending on the involutions on $\mathbb{D}$ and the Clifford algebra. The spinor inner product has the following property
\begin{equation}
(\psi, \phi)=\pm(\phi, \psi)^j
\end{equation}
which are called $\mathbb{D}^j$-symmetric or $\mathbb{D}^j$-skew inner products, respectively. Here $j$ denotes the involution on $\mathbb{D}$. For $\mathbb{D}=\mathbb{R}$, $j$ is the identity, for $\mathbb{D}=\mathbb{C}$, $j$ is the identity or complex conjugation, for $\mathbb{D}=\mathbb{H}$, $j$ is quaternionic conjugation or reversion. Additionally, there is one more option for the choice of inner product on $S$ that comes from the involution on the Clifford algebra. For a Clifford form $\omega$, the spinor inner product satisfies
\begin{equation}
(\psi, \omega.\phi)=(\omega^{\mathcal{J}}.\psi, \phi)
\end{equation}
where $\mathcal{J}$ corresponds to $\xi$ or $\xi\eta$ involutions on the Clifford algebra. $\xi$ acts on any $p$-form $\omega$ as $\omega^{\xi}=(-1)^{\lfloor p/2\rfloor}\omega$ and $\eta$ acts as $\omega^{\eta}=(-1)^p\omega$ where $\lfloor \rfloor$ denotes the floor function. As a result of these properties, one can classify the possible choices of spinor inner products in different dimensions and signatures. The complete classification tables can be found in \cite{Benn Tucker, Ertem Sutemen Acik Catalkaya}.

\section{Gamma matrix notation}

We give the transformation rules between the Clifford calculus and gamma matrix notations in this appendix. Gamma matrices correspond to the representations of Clifford algebra basis which satisfy the following identity
\begin{equation}
\gamma^{\mu}\gamma^{\nu}+\gamma^{\nu}\gamma^{\mu}=2\eta^{\mu\nu}
\end{equation}
where $\eta^{\mu\nu}$ is the flat Lorentzian metric on the background. In a curved background, the coframe basis 1-form can be written in terms of the coordinate basis as $e^a=e^a_{\mu}dx^{\mu}$ where $e^a_{\mu}$ correspond to the tetrad components. From this, we can define the curved space gamma matrices as
\begin{equation}
\gamma^a=e^a_{\mu}\gamma^{\mu}
\end{equation}
and they satify
\begin{equation}
\gamma^a\gamma^b+\gamma^b\gamma^a=2g^{ab}
\end{equation}
where $g^{ab}$ is the inverse metric. So, the Clifford algebra basis $e^a$ in the Clifford calculus notation and the curved space gamma matrices $\gamma^a$ in the gamma matrix notation are identical to each other.

Inner products of spinors are also written in a different form in the gamma matrix notation. For two spinors $\psi$ and $\phi$, the inner product is written as
\begin{equation}
(\psi, \phi)=\overline{\psi}\phi
\end{equation}
and the action of $e_a$ to the second component can be represented as
\begin{equation}
(\psi, e_a.\phi)=\overline{\psi}\gamma_a\phi
\end{equation}
From these principles, we can write all the formulas in the paper in tha gamma matrix from. For example, the Dirac operator is
\begin{equation}
\displaystyle{\not}D=e^a.\nabla_{X_a}=\gamma^a\nabla_a
\end{equation}
where we write the covariant derivative $\nabla_{X_a}=\nabla_a$ in abstract index notation. Similarly, the twistor equation is written as
\begin{equation}
\nabla_a\psi=\frac{1}{n}\gamma_a\displaystyle{\not}D\psi.
\end{equation}
The massless spin-2 field $h_{ab}$ defined in (21) can be written in terms of the Rarita-Schwinger field $\psi_a$ and the twistor spinor $u$ as
\begin{equation}
h_{ab}=\overline{\gamma^cu}\gamma_b\nabla_c\psi_a+\frac{n-2}{n}\overline{\displaystyle{\not}Du}\gamma_b\psi_a+\overline{\psi_a}\gamma_b\displaystyle{\not}Du
\end{equation}
and all other fomulas and calculations in the paper follow the same pattern as given here.

\end{appendix}


\end{document}